# Electrochemical performance of decorated reduced graphene oxide by $MoO_3$ nanoparticles as a counter electrode

Mahyar Servati[1]; Reza Rasuli[1*]

1Department of Physics, Faculty of Science, University of Zanjan,

P.O. Box 45371-38791

**Abstract**

We present an efficient electrocatalytic material based on anchored $MoO_3$ nanoparticles on reduced graphene oxide (RGO) nanosheets. After preparation of graphene oxide (GO), the $MoO_3$ nanoparticles anchored on GO nanosheet by using the arc-discharge method. X-ray diffraction patterns show that the $MoO_3$ nanoparticles are well crystallized on RGO in the orthorhombic crystalline phase with a crystallite size of 83 nm. In addition, FT-IR and Raman spectroscopy results show that during the arc-discharge process, the GO nanosheets have been reduced and RGO nanosheets are decorated with $MoO_3$ nanoparticles which form a porous structure. The surface energy of the prepared electrode was measured as 44.56 $mJ/m^2$, which shows the desirable spreading ability of the electrolyte on the electrode. Finally, electrochemical performance was measured in the symmetrical dummy cell by impedance spectroscopy and cyclic voltammogram, and the photochemical test was measured in the dye-sensitized solar cell by current density measurement. Our results show that the electrochemical performance of the RGM electrode is better than the RGO electrode and is comparable with the Platinum electrode and also the efficiency of RGM electrode used in a dye-sensitized solar cell as a counter electrode is 5.55% near to Platinum electrode performance.

[*] Corresponding author: Tel.: Fax: +98 241 2283203. E-mail address: r_rasuli@znu.ac.ir (Reza Rasuli)

## 1. Introduction

Enhancing the performance of electrochemical and electrocatalytic is important for energy storage and conversion in various devices such as batteries, capacitors, fuel cells, and solar cells [1-3]. In recent years, organic solar cells which are based on electrochemical reactions have attracted remarkable attention of researchers due to the facile and inexpensive production process and an average efficiency of about 12–14% [4-6]. The main effort of the researchers is to reduce production costs, increase efficiency and raise the stability of these type of solar cells [7, 8].

Using organic compounds instead of current costly materials is a challenge towards achieving these three aims [9-12]. In a solar cell, the counter electrode is one of the deterministic components which affects the performance of the cell. For instance, in a dye-sensitized solar cell, by light radiation on the dye molecules, electrons excite and flow into the circuit and then collect via a counter electrode from the external circuit and inject into the electrolyte to reduce dye molecules again [8, 13-15]. In such a system, the counter electrode plays a crucial role to complete this cycle in order to produce a photocurrent. To improve this electrode, researchers have conducted a wide range of their studies on carbonaceous materials for energy conversion and storage, due to its low-cost availability, high chemical stability, proper electrical conductivity as an alternative for platinum [16].

Among carbonaceous materials, graphene has drawn extensive attention as a promising material because of high surface area (2630 $m^2$ $s^{-1}$) and fast charged carrier mobility ( 200,000 $cm^2$ $V^{-1}$ $s^{-1}$) s [17-20]. In electrochemical devices, the number of active sites determines the rate of electrochemical reactions and this site typically located in defects and crystal edges [21]. Enhancing defects and functional groups in graphene allows us to increase the active sites for improving electrocatalytic activity. These factors can be controlled by heating the graphene oxide (GO), and results show that by proper heat treatment of GO, the electron transfer resistance can be reduced to less than 1.20 ohm.$cm^2$ [22]. On the other hand, reducing graphene oxide (RGO) causes eliminate the functional groups in graphene sheets and their active sites in porous structure [23]. Counter electrode can be improved by embedded catalytic nanoparticles between graphene sheets which maintain the electrical conductivity while improving the porous catalytic sites [24].

The $MoO_3$ nanoparticle as an n-type semiconductor with [25] multiple valence states and high chemical and thermal stability, is an efficient material for this purpose, which previously used in organic LED, Li-ion batteries applications and electrocatalyst for Hydrogen Evolution Reaction [26-31]. There is no report on reduced graphene oxide nanosheets functionalized with $MoO_3$ nanoparticles (RGM) as an electrochemical catalyst for $I^-/I_3^-$ redox couple reaction in dye-sensitized solar cells. Triiodide/iodide redox is a common electrolyte in dye-sensitized solar cells and lithium-iodine batteries [12, 32]. Nanostructured $MoO_3$ in proximity to graphene can be used as a counter electrode for application in organic photovoltaic cells. At the interface of graphene and $MoO_3$ nanoparticles, high work function and ionization energy of $MoO_3$ nanoparticles cause to flow down the electrons to $MoO_3$ nanoparticles in order to maintain the thermodynamic equilibrium. This creates a large dipole and increases the hole concentration in the interface of graphene and $MoO_3$ nanoparticles and led to the band bending in $MoO_3$ below to the Fermi level of graphene. As a result, the electron transfer barrier is almost zero and this junction treats similar to a usual metal, and n-type semiconductor junction [33-36]. However, its reaction as an electrocatalyst material with triiodide/iodide redox is not studied yet.

In the following, we will introduce a facile method for synthesis and characterization of RGO nanosheets doped with $MoO_3$ nanoparticles. Then electrochemical test in symmetrical dummy cell will be shown, and the efficiency of the Platinum cathode (Pt) as a reference and RGM electrode obtained by Arc discharge method comparing with RGO electrode which obtained from reduced GO by hydrazine hydrate will be shown in the dye-sensitized solar cell. Electrolyte diffusion in the prepared electrode was explored by the surface energy measurement and finally, the performance of the solar cell made with the prepared counter electrode was investigated by the J-V curve measurement.

## 2. Experimental

*Graphene Oxide synthesis*

The graphene oxide used in this work, synthesized by improved Hummers and Offeman method [37] that have been reported previously [38]. In brief, 15 ml Sulfuric acid added to 1.0 g graphite powder from Sigma Aldrich (45µm powder) and stirred at 27 °C for 15 min. Then In the ice bath, the 3.0 g of potassium permanganate added slowly to the solution. The mixture stirred for 2 h at 40 °C with speed of 1200 RPM. Afterward 300 ml deionized water added to the solution. Using DI water and centrifuge successively, the remaining acid was removed from the solution. To exfoliate graphite oxide to graphene oxide, the solution was exposed in ultrasound waves for 1.5 h and finally was centrifuged to achieve a homogenous dark brown solution.

*RGM synthesis*

To decorate $MoO_3$ nanoparticles on GO nanosheets, we used the arc-discharge method by semi-auto arc discharge device [38, 39]. Two molybdenum electrodes as the cathode and anode with 3 mm diameter and 99.8% purity were placed in GO aqueous solution. Then 10 A currents applied between these two electrodes for 15 minutes. A micrometer screw setting was used to adjust the separation distance between the electrodes. As a result, $MoO_3$ nanoparticles were ablated from the anode and then condensed on the GO solution. The heat generated by the arc-discharge reduced the GO functional groups and stabilized the $MoO_3$ nanoparticles simultaneously and RGM suspension was obtained.

*Symmetrical dummy cell preparation*

To investigate the electrochemical activity symmetrical dummy cell was fabricated from two clean Fluorine-doped Tin Oxide glass (FTO) coated by drop casting method of RGM suspension at 75 °C temperatures. Then they were annealed at a temperature of 400 °C for 2 hours to reduce GO nanosheets completely, crystallizing $MoO_3$ nanoparticles and stabilize material on the FTO glass. These identical cathodes with an active area of 0.1962 $cm^2$ assembled face to face with a 60 μm thick tape as a spacer and sealing. Then the electrolyte (10 ml ethylene glycol, 830 mg potassium iodide, and 127 mg iodine crystals) was injected to cell through a hole on one FTO glass and then closed by a sealing tape.

*Photoelectrochemical cell preparation*

To study on prepared cathode's performance in Dye-sensitized solar cell, Photoelectrochemical cell was fabricated. The $TiO_2$ photoanodes were made of $TiO_2$ paste (9 gr $TiO_2$ powder, 1.5 mL acetic acid, 0.2 mL surfactant ) with the Doctor Blade method on clean FTO glass and then annealed at 325 °C, 375 °C, 450 °C, 500 °C for 5, 5, 15, 15 min respectively to crystallize $TiO_2$ nanoparticles. To sensitize $TiO_2$ nanoparticles, photoanodes were soaked overnight in N719 solution (0.3 mM in a mixed solvent of acetonitrile and tart-butanol with a volume ratio of 1:1). The solar cell was assembled with a dye-sensitized photoanode and RGM, RGO and Pt cathodes, with an active area of 0.1962 $cm^2$. The cathodes were separated by 60 μm thick tape as a spacer and sealing. The electrolyte was injected to cell through the hole on FTO glass, then closed by the tape.

*Characterization*

Fourier transform infrared (FT-IR) spectra were measured with a Thermo Scientific Nicolet iS10 FT-IR spectrometer. The X-ray Diffraction (XRD) experiments were performed using a Philips X-ray diffractometer model PW 1730 by a Cu $k_α$ radiation source with a wavelength of 1.54 Å. Scanning electron microscopy (SEM) and Field Emission Scanning Electron Microscopy (FESEM) were carried out using a TESCAN VEGA-II and TESCAN MIRA-III, respectively. Raman spectroscopy was recorded by Rigaku Handheld Raman Analyzer model FirstGuard with an excitation wavelength of 1064 nm. Electrochemical measurements were performed by using the OGFOIA and OGFEIS multi-channel system of Electrochem

Origalys Company of symmetrical dummy cells. Impedance spectra were measured at 0 V bias voltage and the modulation amplitude of 10 mV, in the frequency range from 65 kHz to 100 mHz. The data of impedance spectra were fitted using Zplot/Zview software. The CV curves were measured at a scan rate of 10 mV/s in 600 to -600 mV voltage range. J-V curves were measured at AM 1.5 G with a radiation power of 100 mW/cm$^2$ by IV100 Cyclic Voltammetry from Safir Soraya Sepahan Company.

## 3. Results and discussion

Using Fourier transform spectroscopy, we studied the reduction process of GO. Figure 1 shows the comparison between the synthesized GO, reduced GO nanosheets by hydrazine hydrate, and RGM obtained by arc discharge method. By adding hydrazine hydrate to the GO, as shown in Figure 1, the functional groups are well reduced. The peak at 3415 cm$^{-1}$ is the characteristic of the O–H bonds due to adsorbed water molecules. The C-H vibrations appear at 2922 cm$^{-1}$ and the peaks at 1730 and 1398 cm$^{-1}$ are correspond to C=O and C-O bonds, respectively. As shown in Figure 1 these peaks have completely disappeared in the RGO and RGM. The peak at 1622 cm$^{-1}$ corresponds to the carbonic bond in graphene, which is observed for all spectra. The C-O-C bond appears at 1077 cm$^{-1}$ and its intensity slightly decreases in both reduction methods [40, 41]. The 1000–400 cm$^{-1}$ peak in RGM corresponds to the vibrations of metal-oxygen characteristic bonds. The peak at 615 cm$^{-1}$ is oxygen bonding with three molybdenum atoms. The peaks at 991 and 877 cm$^{-1}$ are due to the formation of Mo$^{6+}$ atoms bonding with oxygen in the Mo=O and Mo-O-Mo forms, respectively showing crystallization in the orthorhombic phase [25].

To investigate surface morphology, the FESEM images were taken from the prepared GO and RGM powder. As seen in Figure 2, GO plates are well exfoliated and we can see MoO$_3$ nanoparticles with the average size of about 20 nm were anchored uniformly on RGO curved nanosheets.

In Figure 3, the X-ray diffraction spectra of the annealed MoO$_3$ and RGM samples have been presented. In MoO$_3$, the highest intensity of the X-ray diffraction spectrum is due to the (021) plane, which indicates the crystallization of MoO$_3$ in the orthorhombic crystalline phase (α-MoO$_3$) with the spatial group of Pbnm (a = 3.96 Å, b = 13.85 Å, c = 3.69 Å). To calculate the crystallite size, we used the Scherrer's equation as follow:

$$\tau = \frac{K\lambda}{\beta \cos(\theta)}, \tag{1}$$

Where $K$ is the shape factor (= 0.9), $\lambda$ is the X-ray wavelength (= 0.154 nm), $\beta$ is the line broadening at the half maximum intensity and $\theta$ is the Bragg angle. According to XRD results, the crystallite size of the MoO$_3$ obtained about 83 nm. In the RGM spectrum, in addition to the MoO$_3$ peaks, three sharp diffraction peaks are seen at 12.94º, 25.85º and 39.15º, which related to the GO and RGO nanosheets [42-44]. By using Bragg's diffraction law, the spacing of the nanosheets was estimated and the distance between nanosheets for a peak at 12.94º and 25.85º is 6.83, and 3.44 Å, respectively. The high intensity of the RGO peak indicates a significant reduction of GO. The peak at 39.1º with a spacing of nanosheets is equal to 2.3 Å is probably due to a short range order and incomplete oxidation cause them to accumulate [42]. The average crystallite size of the RGM by Scherrer's equation is calculated about 331 nm which shows anchoring of the $MoO_3$ nanoparticles on the RGO.

Raman spectroscopy was used to investigate the structure of the prepared samples (Figure 4). In a typical Raman spectrum of GO, the D peak at 1290 cm$^{-1}$ is breathing mode which is related to the defects and disorders of GO and the G peak at 1599 cm$^{-1}$ represents the graphitized structure and coplanar vibration of sp$^2$-bonded carbon atoms in the two-dimensional hexagonal lattice [45]. In comparison with GO, the intensity of D band relative to G band in the RGM spectrum increases to 1.22, which shows more defects in the structure and high porosity of RGM [46, 47]. According to the empirical equation $L_a$ (nm) = $(2.4 \times 10^{-10}) \lambda^4$ ($I_G/I_D$), the crystallite size of the graphitic carbon is about 252 nm that λ is laser excitation wavelength (1064 nm) [48]. In addition, by comparing the G peak for GO and RGM, the peak has a blue shift indicating the strong bond of nanoparticles to the RGO nanosheets. The peaks at 816 and 976 cm$^{-1}$ are related to the symmetrical and asymmetrical Mo=O stretching vibrations. The peaks at 656 cm$^{-1}$ are assigned to the stretching vibration of O-Mo-O bonds and peaks below 600 cm$^{-1}$ are related to the bending vibrations of the α-$MoO_3$ crystal [46, 49, 50].

Morphologies, surface porosity and cross thickness of the RGM electrode were examined by scanning electron microscopy. Figure 5 (a) shows the SEM images from the anchored $MoO_3$ nanoparticles on the RGO sheets. As shown in Figure 5 (b), after annealing at 400 °C, the nanoparticles are crystallized in a larger size with proper porosity and formed in a columnar shape. Figure 5 (c) exhibit the cross-section of the electrode which shows a layer with a thickness of about 4.5 μm.

In this part, we study the diffusion rate of electrolyte in the RGM electrode. For this purpose, the surface energy by Neumann method was measured. Using the contact angle of five deferent fluids on the electrode surface (as presented in Table 1) and Neumann's equation of state we obtained the surface energy according to the Eq. 2 [51]:

$$cos(\vartheta) = -1 + 2\sqrt{\frac{\gamma_S}{\gamma_L}} e^{-\beta(\gamma_S - \gamma_L)^2} \qquad (2)$$

In Eq. 2, $\vartheta$ is the contact angle of the applied liquid, $\gamma_S$ is the surface energy of the substrate, $\gamma_L$ is the surface energy of the liquid and $\beta$ is a constant. According to the fitted curve in Figure 6 and data in Table 1, the surface energy of RGM was calculated as 44.56 mJ/m$^2$. These results indicate that the RGM electrode is suitable for diffusion of organic electrolytes which can rapidly spread over the electrode surface where ions can react with the electrons.

To characterize the electrochemical activity, the electrochemical impedance measurements and cyclic voltammogram were carried out in a symmetrical dummy cell. By this experiment, the electrochemical performances parameter such as charge transfer resistance at the cathode, electrolyte interface ($R_{CT}$), and overall cell resistance ($R_S$), can be obtained. In Figure 7 the Nyquist plot of electrodes fabricated with Pt, RGO, and RGM are presented. The left inset in this figure shows the magnified curve in the high-frequency region and the right inset indicates an equivalent circuit for fitting Nyquist plot. In this circuit, $R_S$ is the overall Ohmic serial cell resistance. $R_{CT}$ is the charge transfer resistance and CPE is the constant phase element arises from the roughness of the electrodes where describes the deviation from ideal capacitance. $Z_w$ is Nernst diffusion impedance of the triiodide/iodide redox couple in the electrolyte and $Z_{w,pore}$ (that introduced by Mayhew *et al*. [9]) for porous carbon electrodes, is the Nernst diffusion impedance in the porous materials where occurs at high-frequencies.

As we can see in the left inset of Figure 7, Pt acts as a catalytic non-porous surface and the $Z_{w,pore}$ parameter did not appear. So that the high-frequency intercept on the real axis is attributed to the serial resistance ($R_S$) and the first semicircle in the high-frequency region is evaluated as the $R_{CT}$ and the corresponding CPE at the electrode and electrolyte interface. The second semicircle in the low-frequency range arises from Nernst diffusion of electrolyte species ($Z_w$).

The Nyquist plot of RGO and RGM in Figure 7, due to their porous structure, exhibit the high-frequency semicircle that resulting from diffusion through the electrode pores ($Z_{w,pore}$). The slight increase in the $Z_{w,pore}$ of RGM compared with RGO is probably due to the increase in porosity, thus resists against the electrolyte species diffusion in porous structures. In the higher frequency region, the offset value of the real axes indicates the $R_S$ of each sample. In the middle-frequency region, second large semicircle in the RGO and RGM Nyquist plot is determined as $R_{CT}$ and corresponding CPE at the interface of electrode and electrolyte. The $R_{CT}$ of RGM is extremely decrease compared to RGO. This is due to an increase in the surface area of RGM compared to RGO, therefore enhance the electrocatalyst activity. These results are summarized in Table 2. The third semicircle in the low-frequency region due to diffusion impedance of the triiodide/iodide redox couple in the electrolyte is not observed probably because of overlapping with the second semicircle.

We also carried out the cyclic voltammogram test from symmetrical dummy cells with Pt, RGM, and RGO counter electrode. The voltage range was applied between -600 and 600 mV with a scan rate of 10 mV/s. As shown in Figure 8, the maximum current is related to the Pt, RGM, and RGO, respectively. Also, by calculating the slop inverse of the voltammogram curve at zero voltage, it can be approximated to the overall cell resistance [52, 53]. The values for Pt, RGM, and RGO are equal to 0.25, 0.36 and 0.41 Ω respectively.

To characterization the performance of the RGM electrode in an applicable device, we used it in a dye-sensitized solar cell and measured the J-V curve as presented in Figure 9. The photovoltaic parameters indicated in Table 2. The open circuit voltage ($V_{OC}$) is the maximum voltage generated by a solar cell. As shown in the schematic of energy levels in Figure 10, in a dye-sensitized solar cells, $V_{OC}$ is the voltage difference between the oxidation in photoanode and reduction in RGM electrode of the dye molecules. The short-circuit current ($I_{SC}$) is the current through the solar cell when the voltage across the solar cell is zero. By decreasing the overall series resistance of the cell, the $I_{SC}$ increases. Fill Factor (FF) is an interpreter of a deviation from an ideal solar cell that defined as the ratio of the maximum power from the solar cell to the product of $V_{OC}$ and $I_{SC}$. A solar cell with less $R_{CT}$ and $R_S$ shows better FF which is attributed to the increased electrocatalytic performance of cathode materials.

To compare the performance of RGM electrode we used the Platinum electrode as a reference and RGO cathode as a base carbon material, with the same photoanode. According to Table 2, the open-circuit voltage ($V_{OC}$) of the RGM electrode enhances compared with platinum and RGO electrodes. Increasing the porosity in the RGM electrode provides a sufficient number of catalytic sites for electrolyte ions to reduce. $R_S$ reduction leads to increasing $I_{SC}$ of Pt and RGM electrodes in comparison with RGO. As we expected, FF enhanced due to the decrease of $R_{CT}$ in RGM electrode compare with RGO and these results lead to the close efficiency of the RGM electrode to the platinum electrode.

**Conclusions**

In this study, we introduced decorated RGO with $MoO_3$ nanoparticles as a cost-effective counter electrode with good chemical stability and catalytic performance for $I_3^-$ reduction. We presented a facile method to anchor the $MoO_3$ nanoparticles on the GO nanosheets using the arc-discharge method (Figure 11) and our results exhibit that anchored $MoO_3$ nanoparticles are well crystallized on RGO. In addition, FTIR and Raman spectra confirm that the GO reduces and the $MoO_3$ nanoparticles anchored as well on RGO during the arc-discharge process. The low surface energy of the counter electrode shows easy spreading and diffusion of electrolytes in it. Finally, the characterization of electrochemical activity by impedance measurements and cyclic voltammogram results showed that incorporation of $MoO_3$ nanoparticles with RGO nanosheets improve porous structure without degradation electrical conductivity of RGO, so that have a significant impact on the electrochemical efficiency. According to some previous reports of organic materials used in the counter electrode in dye-sensitized solar cell which shows in Table 3, the performance of the dye-sensitized solar cell assembled by RGM electrode indicates that this low-cost electrode can be an appropriate alternative to the Pt electrode in electrochemical cells with acceptable efficiency.

**Acknowledgment**

This work has been supported by Iran National Science Federation (ISNF) under Grant No. 94809262. The authors also thank Dr. Hasan Shayani-Jam, Pouria Moradi for their kind assistance.

Table 1: The surface tension and the contact angle of five different fluids.

| liquid | $\gamma_L$ (mJ/m$^2$) | CA (degree) |
|---|---|---|
| Water | 72.8 | 95 |
| Ethylene Glycol | 47.99 | 64.2 |
| 1-Methyl-2-Pyrrolidinone | 40.79 | 20 |
| n-Hexane | 18.4 | 0 |
| Acetone | 25.2 | 0 |

Table 2: Photovoltaic parameters of RGM and Pt counter electrodes

| Parameters | $I_{SC}$ (mA) | $V_{OC}$ (mV) | FF (%) | Efficiency (%) | $R_S$ ($\Omega$cm$^2$) | $R_{CT}$ ($\Omega$cm$^2$) |
|---|---|---|---|---|---|---|
| RGO | 2.16 | 708 | 55.4 | 4.32 | 8.41 | 124.43 |
| RGM | 2.3 | 752 | 63 | 5.55 | 5.88 | 32.48 |
| Pt | 2.27 | 696 | 72.4 | 5.83 | 5.22 | 6.11 |

Table 3: Electrocatalytic properties of various carbon Counter Electrodes in Dye sensitized solar cells

| Counter Electrode | $J_{SC}$ (mA/cm$^2$) | $V_{OC}$ (mV) | FF (%) | Efficiency (%) |
|---|---|---|---|---|
| functionalized graphene sheets [9] | 7.77 | 710 | 70 | 3.83 |
| Graphene Nanosheets [54] | 13.42 | 690 | 71 | 6.45 |
| Nitrogen-doped porous carbon [55] | 15.58 | 702 | 63 | 7.01 |
| MoS$_2$/Graphene [56] | 17.24 | 714 | 63.83 | 7.86 |
| Mesoporous carbon [57] | 14.32 | 733 | 58 | 6.06 |
| Sub-micrometer-sized Graphite [58] | 12.7 | 794 | 62 | 6.2 |
| Nickel selenide/reduced graphene oxide [59] | 19.94 | 751 | 65 | 9.75 |
| Current work (RGM) | 11.72 | 752 | 63 | 5.55 |

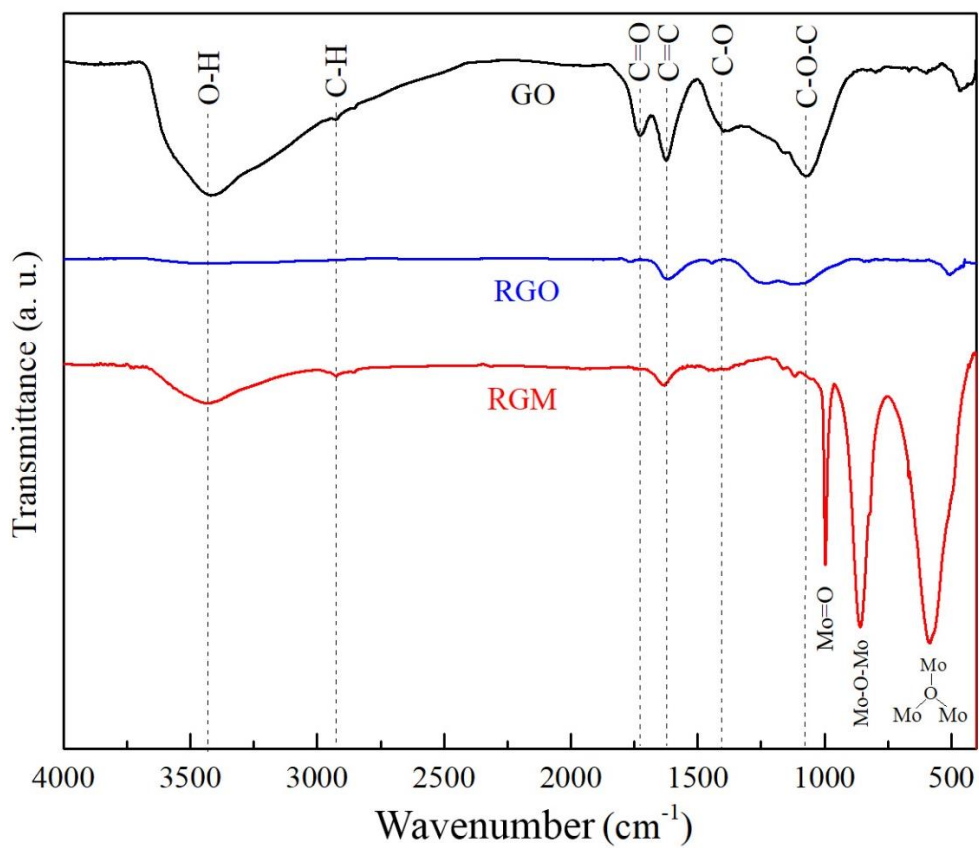

**Figure 1: FTIR analysis of GO and RGO Compare to RGM**

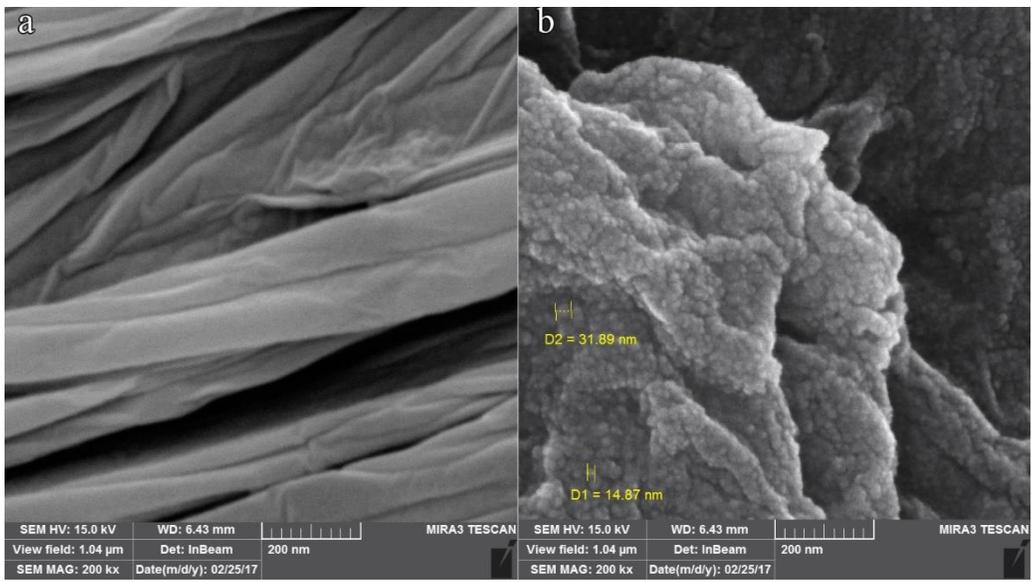

**Figure 2: FESEM images of (a) GO nanosheets and (b) RGM.**

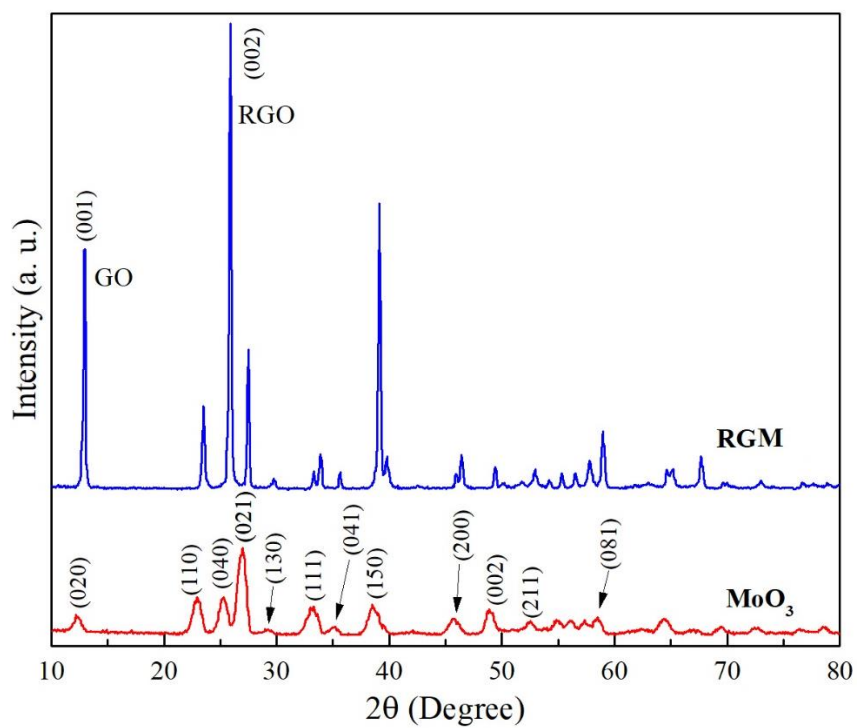

**Figure 3: XRD patterns of MoO3 and RGM, which indicates the crystallization of $MoO_3$ in the orthorhombic crystalline phase.**

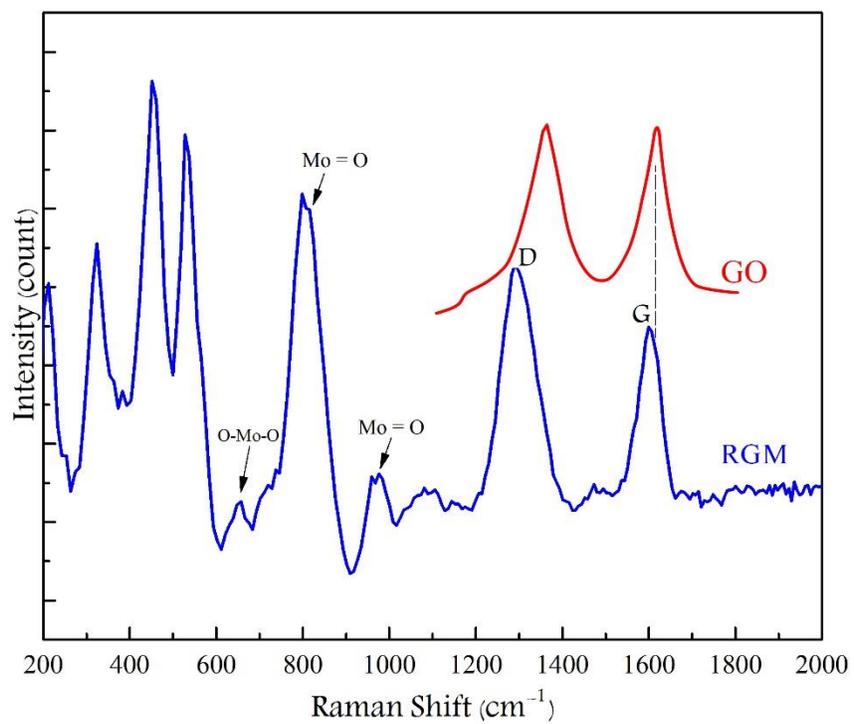

**Figure 4: Raman shift of RGM compared with GO.**

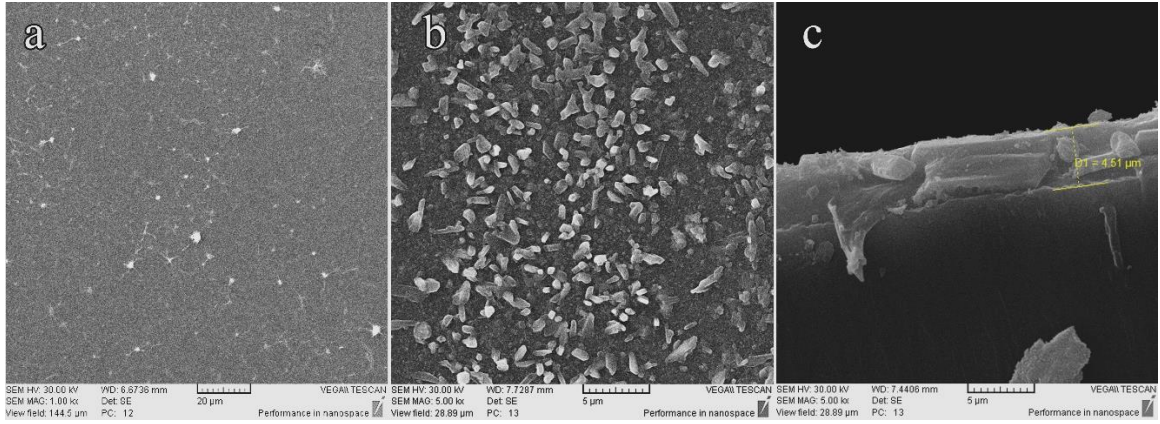

**Figure 5:** FESEM images of electrode surface deposited with RGM (a) before, (b) after annealing at 400 °C and (c) cross-section of the electrode.

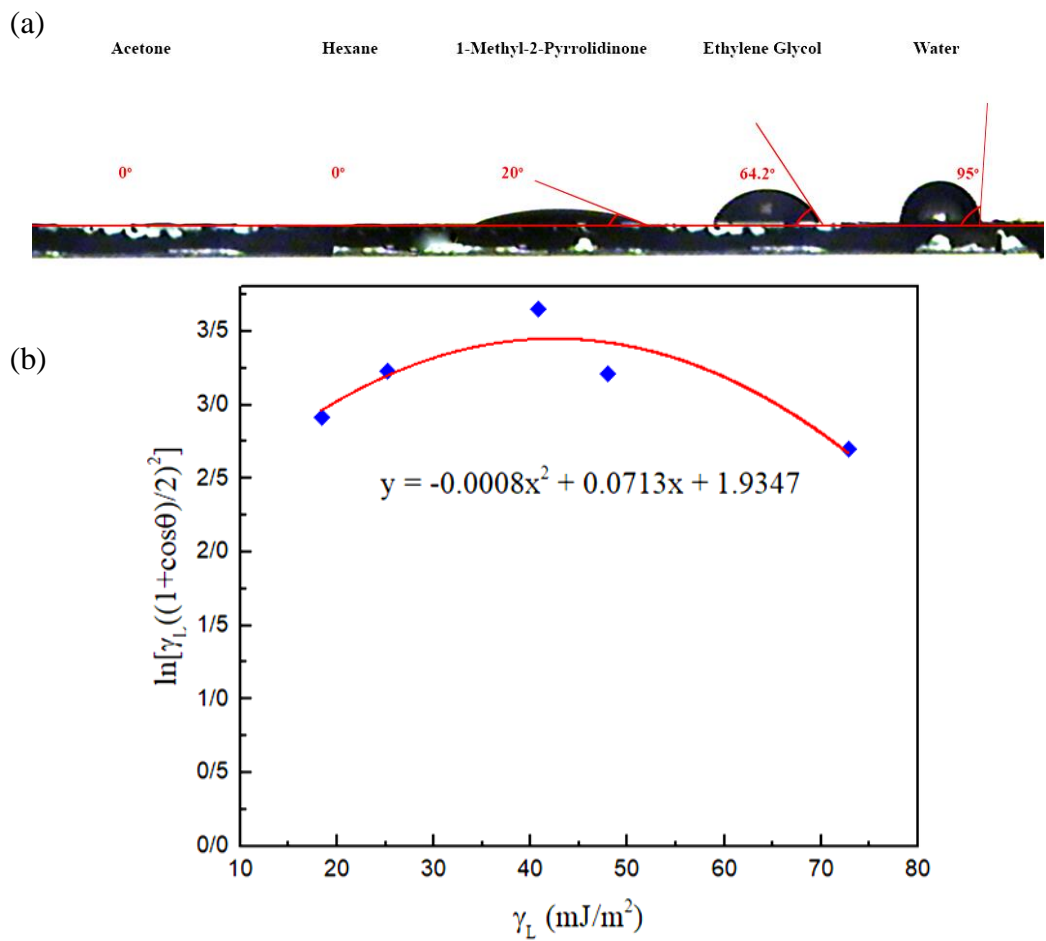

**Figure 6: (a) Images of five various liquid droplets on the RGM surface and (b) the contact angle of each of them. Neumann's equation of state curve.**

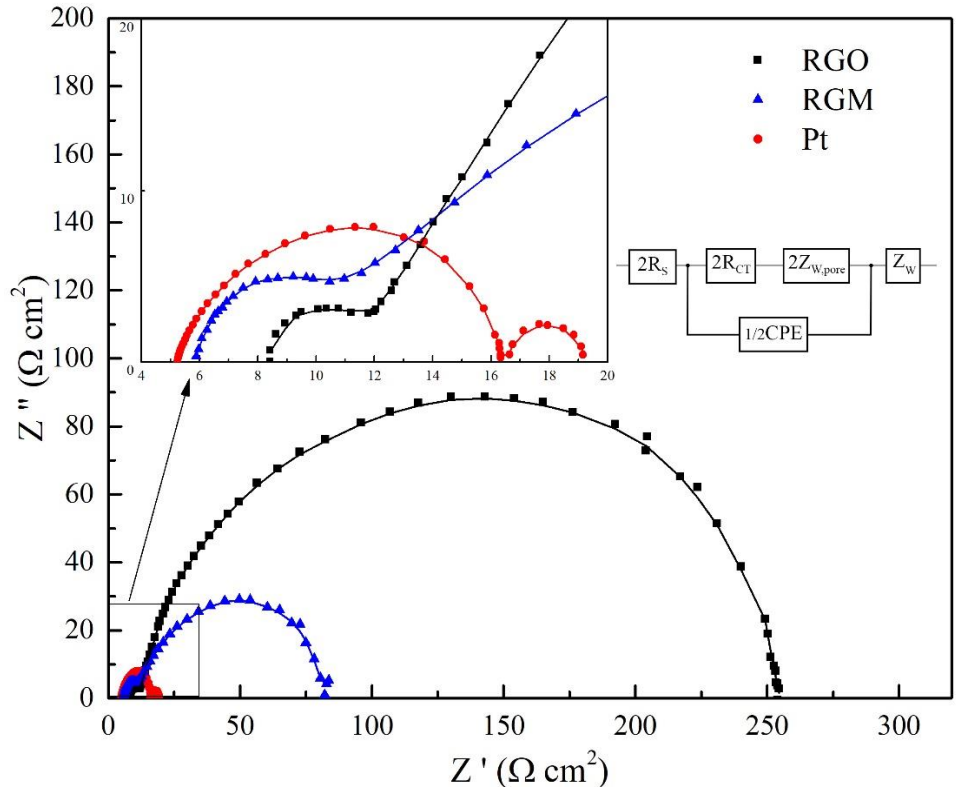

**Figure 7:** The Nyquist plots of symmetrical dummy cells fabricated with RGO, RGM and Pt. Left inset: magnified curves in the high-frequency region. Right inset: Equivalent circuit diagram for fitting the impedance spectra of a dummy cell.

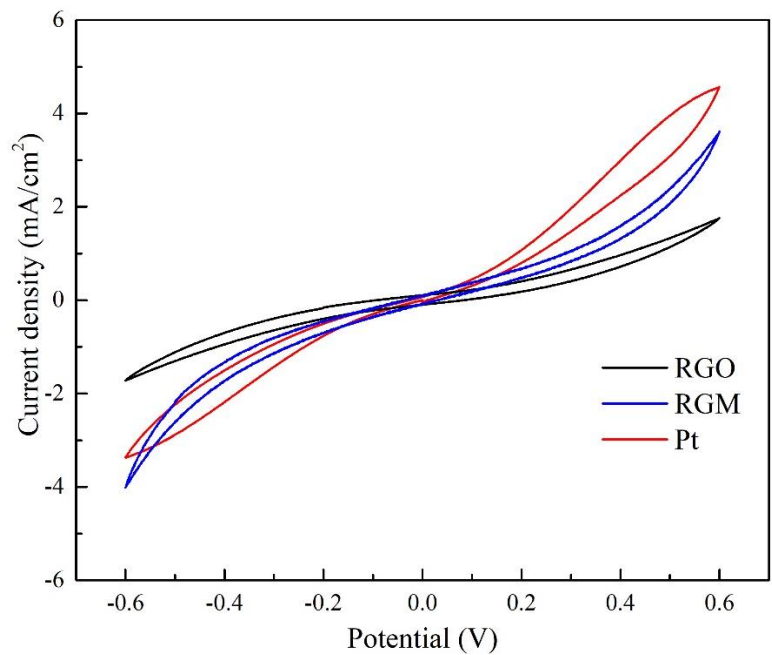

**Figure 8: Cyclic Voltammogram plot of symmetrical dummy cells for RGO, RGM and Pt.**

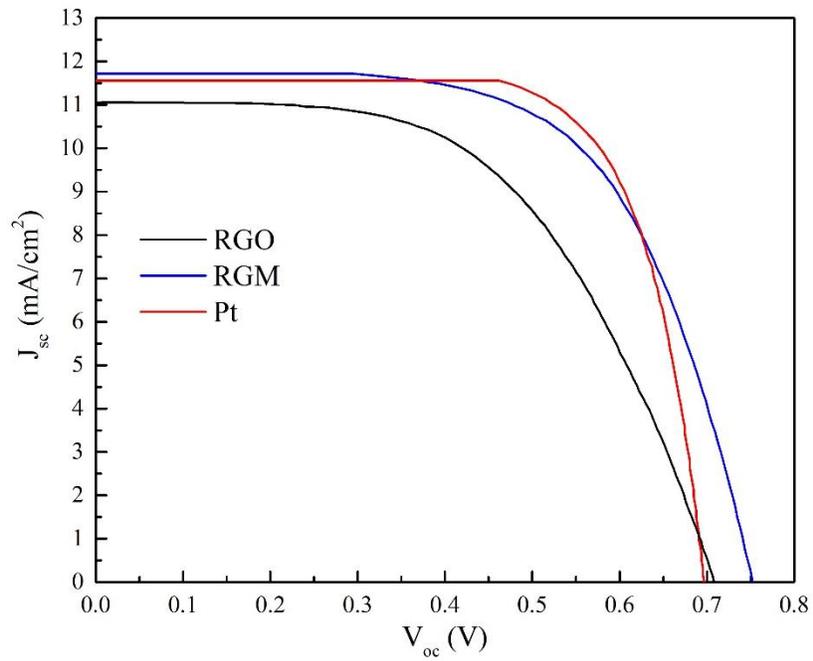

**Figure 9: J-V plot of Photoelectrochemical cell of RGO, RGM, and Pt as a counter electrodes with the identical photoanode.**

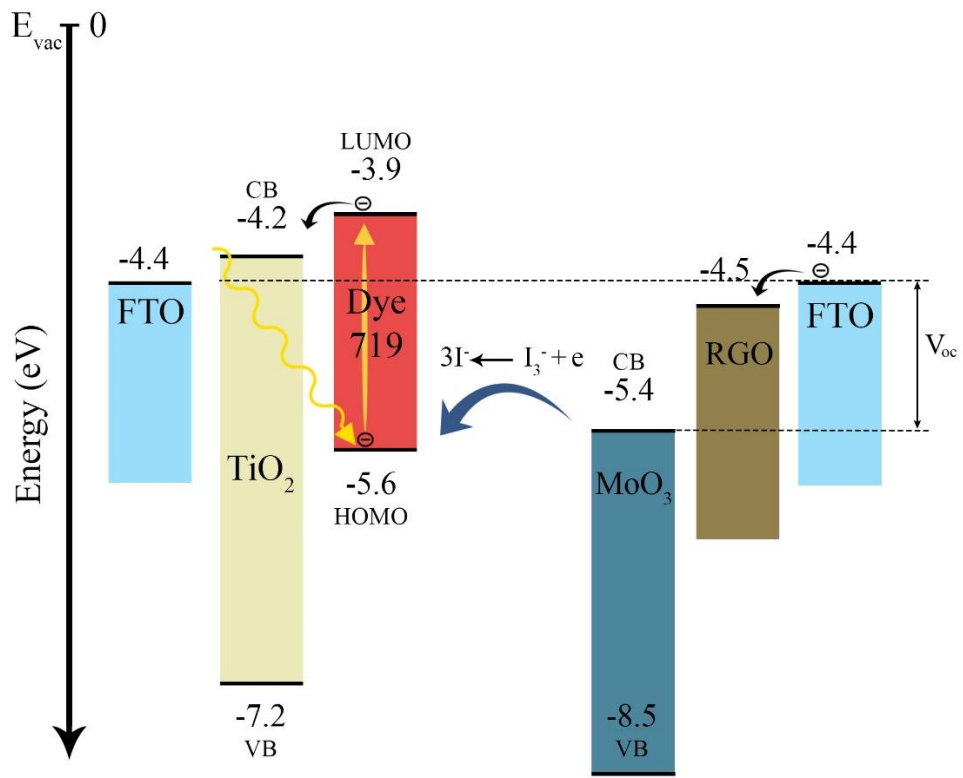

**Figure 10: Schematic of energy levels in different layers of a photovoltaic cell [60]**

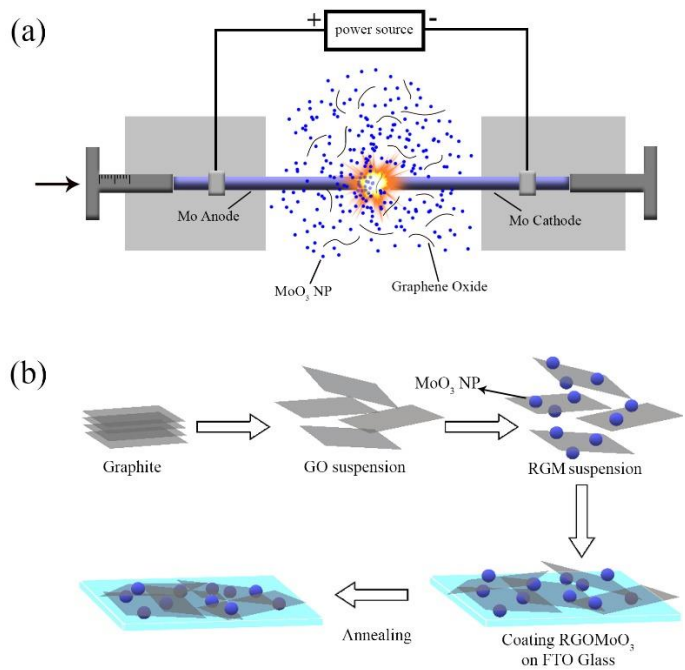

**Figure 11: (a) Schematic of the arc discharge set-up, (b) Schematic illustration of the synthesis and coating process of the cathode.**